\documentstyle[12pt,fleqn]{article}
\begin{document}

\baselineskip 18pt

\begin{flushright}
NEAP-52, December 1995

hep-ph/9512332
\end{flushright}

\vspace*{.5cm}

\begin{center}
{\Large{\bf
Neutron electric dipole moment}}

\bigskip
{\Large{\bf in a supersymmetric model with singlet quarks}}

\vspace{1cm}

{\large{Yoshiya Takeda$ {}^1 $, Isao Umemura$ {}^2 $,
Katsuji Yamamoto$ {}^3 $, Dai Yamazaki}}

\vspace{1cm}

{\it Department of Nuclear Engineering, Kyoto University,
Kyoto 606-01, Japan}

\vspace{1cm}
{\bf abstract}

\end{center}

\bigskip
The neutron electric dipole moment is investigated
in a supersymmetric electroweak model admitting
$ {\rm SU(2)}_W $ singlet and $ Q = -1/3 $, say $ D $-type, quarks.
Significant $ CP $ violating phases appear in the coupling
between the singlet $ D $-type quarks and the ordinary $ d $-type quarks.
Then, through the $ d $-$ D $ mixing effects
on the squark mass terms the gluino one-loop diagrams
provide naturally the contributions $ \sim 10^{-26} e~{\rm cm} $
comparable to the current experimental bound
on the neutron electric dipole moment.
The $ CP $ violation in the $ d $-$ D $ coupling
would also be relevant for the electroweak baryogenesis.

\vspace*{1cm}
\begin{flushleft}
$ {}^1 $ Present address: Hokkaido Electric Power Co., Inc.,
Sapporo 060-91, Japan.

$ {}^2 $ Present address: Taiho 2~-~4~-~4~-~303,
Atsuta-ku, Nagoya 456, Japan.

$ {}^3 $ e-mail address: yamak@yukawa.kyoto-u.ac.jp
\end{flushleft}

\vspace*{1cm}
\begin{flushleft}

keywords: neutron electric dipole moment, supersymmetry, singlet quark

\end{flushleft}

\newpage
The $ CP $ violation is one of the most interesting issues in particle
physics and cosmology.
Among various $ CP $ violating effects the electric dipole moments (EDMs)
of neutron and electron seem to be promissing observables in searching
for new physics, since they are calculated in the standard model
to be much smaller than the values which can be
reached in forthcoming experiments \cite{CP}.
In the minimal supersymmetric standard model (MSSM)
new $ CP $ violating phases arise in the soft supersymmetry breakings
terms.  The EDM of neutron is estimated generally in the MSSM
to exceed the experimental bound \cite{Smith...}
unless the $ CP $ violating phases
in the soft terms are smaller than $ O( 10^{-3} ) $
or the squarks are heavier than a few TeV
\cite{Ellis...,Kizukuri...,Garisto}.
It, however, may be required in the viewpoint of naturalness
that the soft supersymmetry breaking terms
are real and universal at certain unification scale,
if the constraints from the EDMs as well as the flavor-changing processes
\cite{Hagelin...} are considered.
Then, the $ CP $ violating phases in the soft terms
are induced solely from the Cabibbo-Kobayashi-Maskawa (CKM) phase
through the renormalization group evolution.
They are estimated to be rather small, giving the neutron EDM less than
$ 10^{-27}e~{\rm cm} $ \cite{Bertolini...}.
The $ CP $ violation in supersymmetry is also desired
for the electroweak baryogenesis \cite{Cohen...1}.
The parameter ranges to produce the enough amount of baryon
number asymmetry are so restricted in the MSSM
that some strong bounds are placed on the electroweak observables, e.g.
the neutron EDM must be greater than $ 10^{-27}e~{\rm cm} $
\cite{Cohen...2}.
This would rather indicate that the $ CP $ violating phases in the soft
terms induced by the CKM phase through the renormalization group evolution
are not sufficient for the baryogenesis.
The effect of large top Yukawa coupling above the unification scale
can violate significantly the universality of the soft terms.
Then, the CKM-like matter mixings give nontrivial effects
on the soft scalar mass terms at the unification scale,
so that the EDMs of neutron and electron can be as large as
$ 10^{-26} e~{\rm cm} $ \cite{Dimopoulos...}.
These effects might also be relevant for the baryogenesis.

By noting these indistinct aspects on the $ CP $ violation
and also other various motivations,
it will be interesting to consider some extensions of the standard model
and its minimal supersymmetric version,
where new sources of $ CP $ violation are expected to be provided.
We investigate in this letter a supersymmetric electroweak model
which admits $ {\rm SU(2)}_W $ singlet and $ Q = -1/3 $, say $ D $-type,
quark superfields.  This sort of models would be suggested
from the string theory \cite{StringModel}.
New $ CP $ violating phases indeed appear in the coupling between
the singlet $ D $-type quarks and the ordinary $ d $-type quarks.
We examine the $ CP $ violating effects, especially the neutron EDM,
where significant contributions are obtained from the $ d $-$ D $
coupling.

The superpotential relevant for the quark superfields
(suppressing the flavor indicies) is given in a certain basis
denoted with ``0" by
\begin{eqnarray}
\lefteqn{
W = u^c \lambda_u V_0 q_0 H_2
+ d_0^c \lambda_d q_0 H_1 + D_0^c \lambda_\beta q_0 H_1
} \nonumber \\
& ~~~ & \hspace{-1.2em}
+ ~ d_0^c \lambda_\alpha D_0 N + D_0^c \lambda_D D_0 N
+ \lambda_N N H_1 H_2 ~,
\label{eqn:W}
\end{eqnarray}
where $ q_0 = \left( \begin{array}{c} V_0^\dagger u \\
d_0 \end{array} \right) $ with a unitary matrix $ V_0 $,
and the up-type quark mass eigenstates $ u $ and $ u^c $ are taken
with real and diagonal $ \lambda_u $.
It is relevant here to specify the basis without loss of generalty.
The $ \lambda_\beta $ coupling can be rotated out
by a redefinition among the $ d^c_0 $ and $ D^c_0 $ fields.
Then, keeping $ \lambda_\beta = 0 $, it is possible
to diagonalize $ \lambda_d $ and $ \lambda_D $ with real eigenvalues
by unitary transformations for the $ d $ and $ D $ sectors, respectively:
\begin{equation}
\lambda_d = \lambda_{d_i} \delta_{ij} ~,~~
\lambda_D = \lambda_{D_k} \delta_{kl} ~,~~
\lambda_\beta = 0 ~.
\label{eqn:lam}
\end{equation}
In this basis $ V_0 $ is reduced to the CKM matrix $ V $
in the absence of $ d $-$ D $ mixing.
The soft supersymmetry breaking terms associated with eq.(\ref{eqn:W})
are given by substituting the coupling parameters as
\begin{equation}
\lambda_P \rightarrow \xi_P ~~( P = u , d , \alpha , \beta , D , N ) ~.
\label{eqn:L_s}
\end{equation}
Although the $ \lambda_\beta $ coupling is rotated out,
the corresponding $ \xi_\beta $ coupling is generally nonvanishing.

New $ CP $ violating phases appear in the $ N_d \times N_D $ matrix
$ \lambda_\alpha $ in the basis of eq.(\ref{eqn:lam}),
where $ N_d $ and $ N_D $ represent the numbers of the $ d $-type
and $ D $-type quarks, respectively.
Although $ \lambda_\alpha $ admits $ N_d \times N_D $ complex phases,
some of them are physically ineffective,
being rotated out by redefining the relevant fields.
(See ref.\cite{Branco...} for related arguments.)
The phase transformations of $ d_0^c $'s exept for the overall one
should first be made suitably together with those of $ d_0 $'s
keeping $ \lambda_d $ invariant,
so that the usual parametrization of the CKM matrix can be
reproduced up to the small corrections due to the $ \lambda_\alpha $
coupling.
The overall phase transformation of $ d_0^c $'s is, on the other hand,
embedded in the symmetry transformation of $ {\rm U(1)}_Y $.
Hence, only the phase transformations of the $ D $-type quarks
can be used to redefine the $ \lambda_\alpha $ coupling.
In fact, under the $ N_D $ independent transformations,
$ D_{0 k} \rightarrow {\rm e}^{i \phi_k} D_{0k} $,
$ D^c_{0 k} \rightarrow {\rm e}^{-i \phi_k} D^c_{0 k} $,
the $ \lambda_\alpha $ coupling is changed
while the other couplings in eq.(\ref{eqn:W}) remain invariant
in the basis of eq.(\ref{eqn:lam}).
Then, the number of physically relevant $ CP $ violating phases
in $ \lambda_\alpha $ is counted as
$ N_d \times N_D - N_D = (N_d - 1)N_D $.
This implies that at least two $ d $-type quarks
should participate to obtain $ CP $ violating effects
from the $ \lambda_\alpha $ coupling.
It should further be remarked that the CKM matrix may not be involved
in the leading contributions to some observable quantities,
as will be seen later for the gluino contributions to the neutron EDM.
Since the phase convention of the CKM matrix is irrelevant in such cases,
the phase transformations of $ d_0^c $'s exept for the overall one
can be used to redefine $ \lambda_\alpha $.
Then, the effective number of the $ CP $ violating phases
in $ \lambda_\alpha $ is reduced to be
$ (N_d - 1)N_D - (N_d - 1) = (N_d - 1)(N_D - 1) $,
so that at least two $ D $-type quarks are also required
to realize the $ CP $ violating effects from the $ \lambda_\alpha $
coupling without the CKM mixing involved.
If the soft supersymmetry breaking terms are real and universal
at certain unification scale,
they do not provide independent $ CP $ violating phases.
The complex phases of the soft terms are rather induced
through the renormalization group evolution.
It would not be appealing from the viewpoint of naturalness
to consider the soft terms with generic complex phases.
They are in any way stringently constrained by the experimental bounds
on the EDMs \cite{Ellis...,Kizukuri...,Garisto}.

The $ \lambda_\alpha $ coupling involving new $ CP $ violating phases
induces the mixing between the ordinary $ d $-type quarks and
the singlet $ D $-type quarks.
Then, the electroweak gauge couplings of the $ d $-type quarks
are modified through this $ d $-$ D $ mixing,
causing various phenomenological effects.
We hence start with investigating the $ d $-$ D $ mixing effects
by somewhat elaborating the analyses given
in refs.\cite{Branco...,Nir...,Barger...}.

The mass matrix of the $ d $-type and $ D $-type quarks
are given for $ ( d^c_0 , D^c_0 ) \otimes ( d_0 , D_0 ) $ by
\begin{equation}
{\cal M}_{\cal D} = \left( \begin{array}{cc}
\lambda_d v_1 & \lambda_\alpha v_N \\ 0 & \lambda_D v_N
\end{array} \right) ~.
\label{eqn:M_D}
\end{equation}
Here the vacuum expectation values of the Higgs fields
are denoted by
\begin{equation}
\langle H_1 \rangle = v_1 ~,~~ \langle H_2 \rangle = v_2 ~,~~
\langle N \rangle = v_N ~.
\label{eqn:vev}
\end{equation}
These vacuum expectation values are taken to be real at the tree-level,
since the soft $ \xi_N N H_1 H_2 $ term can be made real
by phase transformations among the Higgs fields.
The superpotential (\ref{eqn:W}) actually admits an extra symmetry
$ {\rm U(1)}_E $.  This symmetry may be gauged, or broken explicitly
by the mass and cubic terms of the singlet $ N $ field.
It is, however, not essential in the present investigations
on the $ d $-$ D $ coupling effects which option is really chosen
for the $ {\rm U(1)}_E $.
We here simply assume that $ v_N \sim 100{\rm GeV} - 1{\rm TeV} $,
without considering explicitly the possible effects of $ {\rm U(1)}_E $.

The quark mass matrix (\ref{eqn:M_D}) may be diagonalized
in the following steps.
The $ d $-$ D $ mixing term is first rotated out suitably as
\begin{equation}
{\cal V}_{\rm R}^\dagger {\cal M}_{\cal D} {\cal V}_{\rm L}
= \left( \begin{array}{cc}
M_d^\prime & 0 \\ 0 & M_D^\prime
\end{array} \right) ~.
\label{eqn:M_Dp}
\end{equation}
The unitary matrices are given by
\begin{equation}
{\cal V}_\chi = \left( \begin{array}{cc}
V_{d_\chi} & \epsilon_{1 \chi} \\
- \epsilon_{2 \chi}^\dagger & V_{D_\chi}
\end{array} \right) ~~ ( \chi = {\rm L},~{\rm R} ) ~,
\label{eqn:V}
\end{equation}
and the submatrices are calculated at the leading orders as
\begin{equation}
\epsilon_{1{\rm L}} \simeq \epsilon_{2{\rm L}} \simeq
\epsilon_{\rm L} = (v_1/v_N) \lambda_d \lambda_\alpha
\lambda_D^{-2}~,
\label{eqn:epsilonL}
\end{equation}
\begin{equation}
\epsilon_{1{\rm R}} \simeq \epsilon_{2{\rm R}} \simeq
\epsilon_{\rm R} = \lambda_\alpha \lambda_D^{-1} ~,
\label{eqn:epsilonR}
\end{equation}
\begin{equation}
V_{d_\chi}
\simeq {\bf 1} - {\textstyle{\frac{1}{2}}}
\epsilon_\chi \epsilon_\chi^\dagger ~,~~
V_{D_\chi}
\simeq {\bf 1} - {\textstyle{\frac{1}{2}}}
\epsilon_\chi^\dagger \epsilon_\chi ~.
\label{eqn:VdD}
\end{equation}
The block matrices $ M_d^\prime $ and $ M_D^\prime $ are diagonalized
with real and positive eigenvalues, respectively,
by subsequent unitary transformations
\begin{equation}
{\cal V}^\prime_\chi
= \left( \begin{array}{cc} V_{d_\chi}^\prime & 0 \\
0 & V_{D_\chi}^\prime \end{array} \right) ~.
\label{eqn:V-p}
\end{equation}
Then, the mass eigenstates are given in terms of the original states as
\begin{equation}
\left( \begin{array}{c} d \\ D \end{array} \right)
= ( {\cal V}_{\rm L} {\cal V}_{\rm L}^\prime )^\dagger
\left( \begin{array}{c} d_0 \\ D_0 \end{array} \right) ~,~~
\left( \begin{array}{c} d^c \\ D^c \end{array} \right)
= ( {\cal V}_{\rm R} {\cal V}_{\rm R}^\prime )^{\rm T}
\left( \begin{array}{c} d^c_0 \\ D^c_0 \end{array} \right) ~.
\label{eqn:d_I-M}
\end{equation}

The $ d $-$ D $ mixing effects on the $ d $-type quarks
can be seen by calculating $ M_d^\prime $
with eqs.(\ref{eqn:M_Dp}) -- (\ref{eqn:VdD}):
\begin{equation}
M_d^\prime \simeq \left( {\bf 1} - {\textstyle{\frac{1}{2}}}
\epsilon_{\rm R} \epsilon_{\rm R}^\dagger \right)  \lambda_d v_1  ~.
\label{eqn:Mdp}
\end{equation}
($ {\cal V}_{\rm L} $ gives only sub-leading corrections
suppressed by $ \lambda_d^2 $.)
The $ d $-type quark masses are then modified as
\begin{equation}
m_{d_i} \simeq \lambda_{d_i} v_1 \left[ 1 - {\textstyle{\frac{1}{2}}}
( \epsilon_{\rm R} \epsilon_{\rm R}^\dagger )_{ii} \right] ~,
\label{eqn:md}
\end{equation}
and the CKM matrix is given by
\begin{equation}
V = V_0 V_{d_{\rm L}} V_{d_{\rm L}}^\prime
\simeq V_0 \left( {\bf 1} - {\textstyle{\frac{1}{2}}} \epsilon_{\rm L}
\epsilon_{\rm L}^\dagger \right)
V^\prime_{d_{\rm L}} ~.
\label{eqn:CKM}
\end{equation}
The unitary transformation to diagonalize $ M_d^\prime $ is determined as
\begin{equation}
( V_{d_{\rm L}}^\prime )_{ij} \simeq \delta_{ij}
+ (1 - \delta_{ij}) \frac{m_{d_i} m_{d_j}}{m_{d_i}^2 - m_{d_j}^2}
( \epsilon_{\rm R} \epsilon_{\rm R}^\dagger )_{ij} ~.
\label{eqn:VdLp}
\end{equation}
By noting eq.(\ref{eqn:epsilonR}) and the quark mass ratios,
we obtain
\begin{equation}
V_{d_{\rm L}}^\prime - {\bf 1}
\sim \left( \frac{{\bar \lambda}_\alpha}
{{\bar \lambda}_D} \right)^2 \times
\left( \begin{array}{lll}
{}~ 0     & 10^{-1} & 10^{-3} \\
10^{-1} & ~ 0     & 10^{-2} \\
10^{-3} & 10^{-2} & ~ 0
\end{array} \right) ~,
\label{eqn:VpL-num}
\end{equation}
where $ {\bar \lambda}_\alpha $ and $ {\bar \lambda}_D $ represent
the mean magnitudes of $ \lambda_\alpha $ and $ \lambda_D $, respectively.
Here $ V_{d_{\rm L}}^\prime $ shows a hierarchical structure
similar to that of the CKM matrix.
Hence, up to the small corrections due to the $ d $-$ D $ mixing,
the unitary matrix $ V_0 $ is taken close enough to the actual CKM matrix.

The CKM unitarity is violated in eq.(\ref{eqn:CKM}) as
\begin{equation}
V^\dagger V - {\bf 1}
\simeq - V_{d_{\rm L}}^{\prime \dagger}
( \epsilon_{\rm L} \epsilon_{\rm L}^\dagger ) V_{d_{\rm L}}^\prime ~,
\label{eqn:CKM1}
\end{equation}
\begin{equation}
V V^\dagger - {\bf 1}
\simeq - V V_{d_{\rm L}}^{\prime \dagger}
( \epsilon_{\rm L} \epsilon_{\rm L}^\dagger )
V_{d_{\rm L}}^\prime V^\dagger ~,
\label{eqn:CKM2}
\end{equation}
where $ V_0 \simeq V V_{d_{\rm L}}^{\prime \dagger} $
is considered at the leading order.
By using eq.(\ref{eqn:epsilonL})
and $ \lambda_d \sim 10^{-4} $, $ \lambda_s \sim 10^{-3} $,
$ \lambda_b \sim 10^{-1} $,
the factor to violate the CKM unitarity is estimated as
\begin{eqnarray}
\lefteqn{
( \epsilon_{\rm L} \epsilon_{\rm L}^\dagger )_{ij}
\sim \lambda_{d_i} \lambda_{d_j} ( \lambda_\alpha )_{ik}
( \lambda_\alpha )_{jk}^* \lambda_{D_k}^{-4}
} \nonumber \\
& ~~~ & \nonumber \\
& ~~~ &
\sim \left( \frac{{\bar \lambda}_\alpha}
{{\bar \lambda}_D^2} \right)^2 \times
\left( \begin{array}{lll}
10^{-8} & 10^{-7} & 10^{-5} \\
10^{-7} & 10^{-6} & 10^{-4} \\
10^{-5} & 10^{-4} & 10^{-2}
\end{array} \right) ~.
\label{eqn:ee+num}
\end{eqnarray}
Then, by noting the hierarchical forms of $ V $, $ V_{d_{\rm L}}^\prime $
and $ \epsilon_{\rm L} \epsilon_{\rm L}^\dagger $,
it is readily found that the violation of CKM unitarity is small enough
to satisfy the experimental bounds \cite{ParticleData}.
The $ d $-$ D $ mixing effect in fact appears at the second order of
$ \lambda_d $ in eq.(\ref{eqn:ee+num}) \cite{Branco...}.
This feature can be understood from the following facts:
(i) In the limit $ \lambda_d \rightarrow 0 $ the left-handed $ d $-$ D $
mixing is no longer made in diagonalizing
$ {\cal M}_{\cal D} $ in eq.(\ref{eqn:M_D}),
so that the CKM unitarity is clearly restored.
(ii) For the sign change $ \lambda_d \rightarrow - \lambda_d $,
$ {\cal M}_{\cal D} $ recovers the original form
by the subsequent sign change $ d_0 \rightarrow - d_0 $.

In relation to the violation of CKM unitarity,
the $ d $-$ D $ mixing also modifies the neutral currents
of the $ d $-type quarks coupled to the $ Z $ boson
\cite{Branco...,Nir...,Barger...}.
By using eqs.(\ref{eqn:V}) -- (\ref{eqn:d_I-M}), the neutral currents
are expressed in terms of the mass eigenstates as
\begin{equation}
{\cal J}^Z_\mu = d^\dagger \sigma_\mu \left( - \frac{x_W}{3} {\bf 1}
+ \frac{z}{2} \right) d + d^{c \dagger} \sigma_\mu
\left( \frac{x_W}{3} {\bf 1} \right) d^c + \ldots ~,
\label{eqn:JZ}
\end{equation}
where $ x_W \equiv \sin^2 \theta_W $.
The modification by the $ d $-$ D $ mixing is given by
\begin{equation}
z - {\bf 1} \simeq V_{d_{\rm L}}^{\prime \dagger}
( \epsilon_{\rm L} \epsilon_{\rm L}^\dagger ) V_{d_{\rm L}}^\prime ~,
\label{eqn:z}
\end{equation}
which has already been observed in eq.(\ref{eqn:CKM1}).
Here, the flavor changing neutral currents indeed arise
with the coupling factors $ z_{ij} $ ($ i \not= j $)
involving the $ CP $ violating phases, while the flavor diagonal ones
are reduced by the small amounts of $ | z_{ii} - 1 | $.

The experimental bounds on the $ d $-$ D $ mixing effects including
$ CP $ violating ones have been investigated extensively
in the $ K_0 $-$ {\bar K}_0 $ mixing, $ B_0 $-$ {\bar B}_0 $
mixing, $ K_L \rightarrow \mu^+ \mu^- $, and so on
\cite{Branco...,Nir...,Barger...}.
Here, by taking the coupling matrices given in the basis
of eq.(\ref{eqn:lam}), it has already been observed
in eqs.(\ref{eqn:VpL-num}) and (\ref{eqn:ee+num})
that the $ d $-$ D $ mixing effects are small enough
by virtue of the hierarchical quark masses and the CKM mixing.
Therefore, the experimental bounds on the neutral currents
and the CKM unitarity do not place stringent constraints
on the $ \lambda_\alpha $ coupling, just requiring
$ {\bar \lambda}_\alpha
{}~{\mbox{~$ < $ \hspace{-1.05em}{\raisebox{-0.75ex}{$ \sim $}}~}}~
{\bar \lambda}_D^2
{}~{\mbox{~$ < $ \hspace{-1.05em}{\raisebox{-0.75ex}{$ \sim $}}~}}~ 1 $.

As seen in the above, it is possible that the $ \lambda_\alpha $ coupling
has $ {\bar \lambda}_\alpha \sim 0.1 - 1 $ with complex phases
of $ O(1) $.  Then, the $ CP $ violating effects by the $ \lambda_\alpha $
coupling will appear significantly in some physical processes.
We here investigate especially the neutron EDM, where novel contributions
are obtained from the $ \lambda_\alpha $ coupling.

It is first noted that the one-loop diagrams involving the gauge bosons
do not contribute even in the presence
of $ d $-$ D $ mixing.  This is understood in the following way.
The neutral currents of the right-handed $ d $-type quarks
coupled to the $ Z $ boson are still flavor-diagonal,
since $ d^c $ and $ D^c $ fields have the same quantum numbers.
Then, if the right-handed coupling to $ Z $ is taken for at least one of
the vertices in a loop diagram, the intermediate quark state
must be the $ d $ quark itself for the $ d $ quark EDM.
The flavor-diagonal left-handed and right-handed
$ {\bar d} $-$ d $-$ Z $ couplings are, however, real due to hermiticity.
On the other hand, if the left-handed coupling is taken for both
the vertices, the complex phases at the two vertices cancel.
Hence, the one-loop diagram involving the $ Z $ boson
does not contribute to the $ d $ quark EDM.
As for the charged gauge couplings, the right-handed quarks
do not coupled to the $ W $ boson.  Then, the external $ d $ quark mass
insertion is made to flip the chirality for contributing to the EDM.
However, the complex phases at the two charged current vertices cancel
in the one-loop diagrams involving the $ W $ boson.
The same arguments are also applied to the $ u $ quark EDM.
These situations are similar to those in the standard model
\cite{CP}.

It may be supposed naively that some contributions to the neutron EDM
are obtained through the $ d $-$ D $ mixing in one-loop diagrams
consisting of the physical charged Higgs boson
$ H^+ = - \sin \beta H_1^{- \dagger} + \cos \beta H_2^+ $
($ \tan \beta = v_2 / v_1 $) and the quark intermediate states.
However, these contributions are vanishing, just as the $ W $ boson
contributions.  In fact, the charged Yukawa couplings to $ H_1^- $
and $ H_2^+ $ are expressed as follows
in terms of the quark mass eigenstates,
so that the right couplings to the Goldstone mode
$ G^+ = \cos \beta H_1^{- \dagger} + \sin \beta H_2^+ $
are reproduced:
\begin{equation}
\lambda_u^{H_2} = \lambda_u V_0 V_{d_{\rm L}} V_{d_{\rm L}}^\prime
= ( M_u / v_2 ) V ~,
\label{eqn:lH2}
\end{equation}
\begin{equation}
\lambda_d^{H_1} = ( V_{d_{\rm R}} V_{d_{\rm R}}^\prime )^\dagger
\lambda_d V_0^\dagger = ( M_d / v_1 ) V^\dagger ~,
\label{eqn:lH1}
\end{equation}
where $ M_u $ ($ = \lambda_u v_2 $) and $ M_d $
($ = V_{d_{\rm R}}^{\prime \dagger} M_d^\prime V_{d_{\rm L}}^\prime $)
are the diagonal quark mass matrices with real and positive eigenvalues.
This shows that the physical $ H^+ $ also has the Yukawa couplings
with the same flavor structure as the charged gauge couplings.
Therefore, the one-loop $ H^+ $ contributions to the neutron EDM
are clearly vanishing.
(This is also checked by a numerical calculation.)
The supersymmetric counter parts of the diagrams, which involve
the charginos instead of the charged Higgs boson, however,
can contribute to the neutron EDM,
if some nontrivial flavor mixings are induced
for the squark mass terms in the super-flavor basis
associated with the quark mass eigenstates.
Such chargino contributions could be comparable
to the gluino contributions, which are examined in detail later.

The one-loop diagrams consisting of the neutral Higgs bosons
and the $ D $-type quarks are, on the other hand, expected
to give some contributions to the $ d $ quark EDM
through the $ d $-$ D $ coupling.
The $ d $-type and $ D $-type quark couplings
to the neutral Higgs fields, however, have the specific forms
as $ p {\cal M}_{\cal D}/v_N + q \Lambda ( \lambda_d ) $
in the basis of eq.(\ref{eqn:lam}),
where the diagonal block of $ \Lambda ( \lambda_d ) $
for the $ d $ sector is the $ N_d \times N_d $ matrix $ \lambda_d $,
while the other blocks of $ \Lambda ( \lambda_d ) $ are vanishing,
and $ ( p , q ) = ( 1 ,  - v_1 / v_N ) $ for $ N $,
$ ( 0 , 1 ) $ for $ H_1^0 $.
($ H_2^0 $ does not have this sort of coupling.)
Then, the $ d $-$ D $ couplings are obtained
as $ O( \lambda_d^2 \lambda_\alpha ) d^c D \Phi $
and $ O( \lambda_d \lambda_\alpha ) D^c d \Phi $
in terms of the quark mass eigenstates and $ \Phi = N, H_1^0 $,
so that the neutral Higgs boson contributions arise at the order
of $ \lambda_d^3 $.
It is, on the other hand, noted that the terms
$ \sim \lambda_d^3 \lambda_\alpha^2 $ do not contribute
to the $ d $ quark EDM, since they should be real as
$ ( \lambda_\alpha )_{ik} ( \lambda_\alpha^\dagger )_{ki}
= | ( \lambda_\alpha )_{ik} |^2 $ being invariant under
the phase transformations of the quark fields
to rearrange the complex phases in $ \lambda_\alpha $.
Therefore, the neutral Higgs boson contributions
through the $ d $-$ D $ coupling are found to be obtained
at the order of $ \lambda_d^3 \lambda_\alpha^4 $,
which could also be comparable to the gluino contributions
presented below.

We now examine in detail the gluino one-loop diagrams for the neutron EDM,
as shown in fig.1, where some significant contributions are expected
to be obtained from the $ d $-$ D $ coupling $ \lambda_\alpha $.
In order to calculate the gluino contributions, it is relevant to find
the squark mass eigenstates.
The mass squared matrix of the $ {\tilde d} $ and $ {\tilde D} $
squarks are given as follows in the basis
$ [ ( {\tilde d}, {\tilde D} ) ,
( {\tilde d}^{c \dagger}, {\tilde D}^{c \dagger} ) ] $
associated with the quark mass eigenstates (\ref{eqn:d_I-M}):
\begin{equation}
{\mbox{\large{\sf M}}}^2_{\tilde{\cal D}}
= \left( \begin{array}{cc}
{\cal M}^2_{\rm L} & m^2 \Delta_{\rm LR}^\dagger \\
{ } & { } \\
m^2 \Delta_{\rm LR} &
{\cal M}^2_{\rm R}
\end{array} \right) ~,
\label{eqn:MMD^2}
\end{equation}
where $ m \sim 10^2 {\rm GeV} $ represents the characteristic
scale of the soft supersymmetry breakings.
The diagonal blocks $ {\cal M}^2_{\rm L} $ and $ {\cal M}^2_{\rm R} $
contain the contributions of the quark masses, gauge $ D $-terms
and soft supersymmetry breaking terms.  The off-diagonal block
to connect the left-handed and right-handed squarks is given
(in the basis with $ \lambda_\beta = 0 $) by
\begin{equation}
\Delta_{\rm LR} = \left( \begin{array}{cc}
\Delta_d & \Delta_\alpha \\
\Delta_\beta & \Delta_D
\end{array} \right)
= ( {\cal V}_{\rm R} {\cal V}_{\rm R}^\prime )^\dagger
\Delta_{\rm LR}^0
( {\cal V}_{\rm L} {\cal V}_{\rm L}^\prime )~,
\label{eqn:DLR}
\end{equation}
\begin{equation}
\Delta_{\rm LR}^0 = m^{-2} \left( \begin{array}{cc}
\xi_d v_1 + \lambda_d \lambda_N v_2 v_N
     & \xi_\alpha v_N + \lambda_\alpha \lambda_N v_1 v_2 \\
\xi_\beta v_1 & \xi_D v_N + \lambda_D \lambda_N v_1 v_2
\end{array} \right) ~.
\label{eqn:DLR0}
\end{equation}

The squark mass eigenstates
$ {\tilde{\cal D}}_K \equiv ( {\tilde d}_{+i} , {\tilde D}_{+k} ,
{\tilde d}_{-j} , {\tilde D}_{-l} ) $ are found by diagonalizing
$ {\mbox{\large{\sf M}}}^2_{\tilde{\cal D}} $
with a suitable unitary transformation
$ {\mbox{\large{\sf U}}}_{\tilde{\cal D}} $,
where the compressed index $ K $ denotes simultaneously the flavors
and the four varieties of mass eigenstates.
Then, the squark states associated with the $ d $-type quark
mass eigenstates are, in particular, given as
\begin{equation}
{\tilde d}_i = \sum_K
({\mbox{\large{\sf U}}}_{\tilde{\cal D}})_{{\tilde d}_i K}
{\tilde{\cal D}}_K ~,~~
{\tilde d}_i^{c \dagger} = \sum_K
({\mbox{\large{\sf U}}}_{\tilde{\cal D}})_{{\tilde d}_i^{c \dagger} K}
{\tilde{\cal D}}_K ~.
\label{eqn:sd}
\end{equation}
By using eq.(\ref{eqn:sd}), the gluino contribution $ d_d ({\tilde g}) $
to the $ d (= d_1) $ quark EDM is calculated as
\begin{equation}
d_d ({\tilde g}) = e \frac{2 \alpha_s}{9 \pi m_{\tilde g}}
\sum_K {\rm Im} \left[
({\mbox{\large{\sf U}}}_{\tilde{\cal D}})_{{\tilde d}_1^{c \dagger} K}
({\mbox{\large{\sf U}}}^\dagger_{\tilde{\cal D}})_{K {\tilde d}_1}
\right] X_K I ( X_K ) ~,
\label{eqn:d_d}
\end{equation}
where $ I( X_K ) $ is some relevant function
of $ X_K \equiv m_{\tilde g}^2 / m_{{\tilde{\cal D}}_K}^2 $
with the gluino mass $ m_{\tilde g} $
(real by taking a suitable phase convention)
and the squark masses $ m_{{\tilde{\cal D}}_K} $.

The unitary matrix $ {\mbox{\large{\sf U}}}_{\tilde{\cal D}} $ may be
determined in the expansion with respect to $ \Delta_{\rm LR} $
(with diagonal $ {\cal M}^2_{\rm L} $
and $ {\cal M}^2_{\rm R} $ for simplicity
though they get some nonvanishing off-diagonal components,
as discussed later).
Then, the gluino contribution is calculated as
\begin{equation}
d_d ({\tilde g}) = e \frac{2 \alpha_s}{9 \pi m_{\tilde g}}
\left\{ h_1 {\rm Im} \left[ ( \Delta_d )_{11} \right]
+ {\cal O} ( \Delta_{\rm LR} \Delta_{\rm LR}^\dagger \Delta_{\rm LR} )
\right\} ~,
\label{eqn:d-d1}
\end{equation}
where $ h_1 \sim 1 $ is a factor depending on $ X_{{\tilde d}_{\pm 1}} $
and $ m_{\tilde g}^2 / m^2 $.
The $ \Delta_d $ term reproduces the usual MSSM contribution
for $ \lambda_\alpha \rightarrow 0 $.
The contributions of
$ {\cal O} ( \Delta_{\rm LR} \Delta_{\rm LR}^\dagger \Delta_{\rm LR} ) $
involve the terms such as
$ ( \Delta_\alpha )_{1k}( \Delta_\alpha^\dagger )_{kj}( \Delta_d )_{j1} $,
$ ( \Delta_\alpha )_{1k}( \Delta_D^\dagger )_{kl}( \Delta_\beta )_{l1} $,
$ ( \Delta_d )_{1i}( \Delta_d^\dagger )_{ij}( \Delta_d )_{j1} $ and
$ ( \Delta_d )_{1i}( \Delta_\beta^\dagger )_{ik}( \Delta_\beta )_{k1} $.
If the gluino contribution is expressed in terms of the coupling matrices
$ \lambda_d $, $ \lambda_\alpha $ and so on, the terms in
$ {\cal O} ( \Delta_{\rm LR} \Delta_{\rm LR}^\dagger \Delta_{\rm LR} ) $
generally provide the contributions at the same orders
as those obtained from the $ \Delta_d $ term.

We are especially interested in the case
where the soft supersymmetry breaking terms have the universal form
at the unification scale $ M_U $
with real gaugino masses and a common real factor $ A $
for the scalar cubic couplings $ \xi_P $.
Then, for illustrating the contributions from the $ \lambda_\alpha $
coupling, let us first consider the universal limit by neglecting
the the renormalization group effects.
In this specific case $ \Delta_{\rm LR}^0 $
(with $ \xi_\beta = Am \lambda_\beta = 0 $) is written
in a rather restricted form as
\begin{equation}
\Delta_{\rm LR}^0 ({\rm universal}) = a \left( \begin{array}{cc}
\lambda_d & 0 \\ 0 & 0 \end{array} \right)
+ ( b / v_N ) {\cal M}_{\cal D} ~,
\label{eqn:DLR0-u}
\end{equation}
where $ a = (v_1 / m )[ A + \lambda_N ( v_2 v_N / m v_1 ) ]
- ( v_1 / v_N )b $
and $ b = (v_N/ m )[ A + \lambda_N ( v_1 v_2 / m v_N ) ] $.
The second term of eq.(\ref{eqn:DLR0-u}) is diagonalized for calculating
$ \Delta_{\rm LR} $ in eq.(\ref{eqn:DLR})
together with the quark mass matirx.
Then, the $ \Delta_d $ term is calculated  as
\begin{equation}
\Delta_d = (v_1 / m )[ A + \lambda_N ( v_2 v_N / m v_1 ) ] \lambda_d
+ \delta \Delta_d
\label{eqn:Dd-u}
\end{equation}
with $ \delta \Delta_d
= a [ ( V_{d_{\rm R}} V_{d_{\rm R}}^\prime )^\dagger
\lambda_d ( V_{d_{\rm L}} V_{d_{\rm L}}^\prime ) - \lambda_d ] $.
The first term in eq.(\ref{eqn:Dd-u}), which is independent
of $ \lambda_\alpha $, reproduces the usual gluino contribution
in the MSSM by identifying $ \lambda_N v_N $ with the so-called
$ \mu $-term (though it is vanishing with the real and universal
soft supersymmetry breaking terms),
while the effects of $ \lambda_\alpha $  appears
in $ {\rm Im} [ ( \delta \Delta_d )_{11} ] $.
It is, however, found by noting eqs.(\ref{eqn:CKM}) and (\ref{eqn:lH1})
that
\begin{equation}
\delta \Delta_d = a [ ( M_d / v_1 ) V^\dagger V - \lambda_d ] ~,
\label{eqn:dDd}
\end{equation}
where $ M_d $ is the diagonal $ d $-type quark mass matrix
with real and positive eigenvalues.
Therefore, the contribution of the $ \Delta_d $ term
in eq.(\ref{eqn:d-d1}) is vanishing with the real $ A $
and real gaugino masses in the universal limit.
(This is checked by a numerical calculation.)

The off-diagonal blocks in $ \Delta_{\rm LR} $ are brought
only from the first term in eq.(\ref{eqn:DLR0-u}) as
$ \Delta_\alpha \sim \lambda_d \epsilon_{\rm L} $
and $ \Delta_\beta \sim \epsilon_{\rm R}^\dagger \lambda_d $.
Then, some contributions to $ d_d ({\tilde g}) $ are obtained
at the order of $ \lambda_d^3 $ in
$ {\cal O} ( \Delta_{\rm LR} \Delta_{\rm LR}^\dagger \Delta_{\rm LR} ) $
of eq.(\ref{eqn:d-d1}).
It should here be remembered that the terms
$ \sim \lambda_d^3 \lambda_\alpha^2 $ do not contribute
to $ d_d ({\tilde g}) $, since they should be real as
$ ( \lambda_\alpha )_{ik} ( \lambda_\alpha^\dagger )_{ki}
= | ( \lambda_\alpha )_{ik} |^2 $ being invariant under
the phase transformations of the quark fields to rearrange
the complex phases in $ \lambda_\alpha $.
Therefore,
the $ {\cal O}( \Delta_{\rm LR} \Delta_{\rm LR}^\dagger \Delta_{\rm LR} ) $
term in eq.(\ref{eqn:d-d1}) provides the gluino contribution
$ d_d ({\tilde g}) $ in the universal limit
with the $ CP $ violating factor as
\begin{equation}
{\bar \lambda}_D^{-4} \lambda_{d_1} \lambda_{d_j}^2
{\rm Im} \left[ ( \lambda_\alpha )_{1k} ( \lambda_\alpha^\dagger )_{kj}
( \lambda_\alpha )_{jl} ( \lambda_\alpha^\dagger )_{l1} \right] ~.
\label{eqn:ImdDd}
\end{equation}
This gluino contribution is then estimated as
\begin{equation}
d_d ({\tilde g}) \sim 10^{-26} e~{\rm cm}
\left( \frac{100{\rm GeV}}{m} \right)
\left( \frac{{\bar \lambda}_\alpha / {\bar \lambda}_D}{0.3} \right)^4
\phi_\alpha ~,
\label{eqn:gnum}
\end{equation}
with $ \lambda_{d_1} = \lambda_d \sim 10^{-4} $,
$ \lambda_{d_j} = \lambda_b \sim 10^{-1} $,
$ m_{{\tilde{\cal D}}_K} , m_{\tilde g} \sim m $,
and the $ CP $ violating phase $ \phi_\alpha $
from the $ \lambda_\alpha $ coupling.
(This estimate has actually been confirmed by a numerical calculation.)
The $ CP $ violating factor in eq.(\ref{eqn:ImdDd}) is clearly expressed
in a form invariant under the relevant phase transformations
of the quark fields.
It is readily found that this factor is vanishing
if $ N_d = 1 $ ($ j = 1 $) or $ N_D = 1 $ ($ k , l = 1 $),
in agreement with the $ CP $ violating phase counting mentioned before
for the cases without the CKM matrix involved.

It should also be noted that even in the universal limit
$ {\cal M}^2_{\rm L} $ gets some off-diagonal elements
from the gauge $ D $-terms through the quark mixings,
since the left-handed $ d_0 $ and $ D_0 $
fields have the different gauge quantum numbers.
(The quark mixings involving the CKM one are, on the other hand,
rotated out in the universal soft squark mass terms.)
The off-diagonal elements of $ {\cal M}^2_{\rm L} $ induce for instance
a $ {\tilde D}_+ $ component
$ \sim \epsilon_{\rm L} \sim \lambda_d \lambda_\alpha $
in the squark state $ {\tilde d}_1 $.
On the other hand, the squark state $ {\tilde d}^{c \dagger}_1 $
has a $ {\tilde D}_+ $ component $ \sim \Delta_\alpha $.
Then, a contribution
$ \sim {\rm Im}[ ( \Delta_\alpha \epsilon_{\rm L}^\dagger )_{11} ]
\sim \lambda_d^3 \lambda_\alpha^4 $ to the $ d $ quark EDM is obtained
with the intermediate $ {\tilde D}_+ $ state.
This $ d $-$ D $ mixing effect on the gauge $ D $-terms
of the squark masses hence provide the gluino contribution
comparable to that given in eq.(\ref{eqn:gnum}).

The gluino contributions at the order of $ \lambda_d \lambda_\alpha^4 $
are rather obtained through the renormalization group effects,
which will be significant with the factor
$ t_U = ( 4 \pi )^{-2} \ln ( M_U / M_W ) \sim 0.2 $
for $ M_U / M_W \sim 10^{14} $.
For example, the deviation from the universality is induced
as $ \xi_\alpha \xi_D^{-1} - 1 \sim t_U {\bar \lambda}_D^2 $,
so that $ \Delta_{\rm LR}^0 $ no longer has the specific form
as given in eq.(\ref{eqn:DLR0-u}).
Then, $ \Delta_\alpha $ becomes of the order of
$ \lambda_\alpha $ rather than $ \lambda_d \epsilon_{\rm L} $,
and $ ( \delta \Delta_d )_{11} $ has the imaginary part
at the first order of $ \lambda_d $.
The $ \xi_\beta $ coupling gets a contribution
$ \sim m ( t_U {\bar \lambda}_D^2 )
\lambda_D^{-1} \lambda_\alpha^\dagger \lambda_d $,
which gives a correction to $ \Delta_\beta $
at same order as the universal limit value.
By taking into account these effects for the $ \Delta_{\rm LR} $ coupling
in eq.(\ref{eqn:d-d1}), we obtain the contributions
to $ d_d ({\tilde g}) $ with the $ CP $ violating factors
$ \sim t_U \lambda_d \lambda_\alpha^4 $, which can be comparable
to the factor given in eq.(\ref{eqn:ImdDd}) for the universal limit.

It should further be noted that some contributions are generated
in the off-diagonal elements of $ {\cal M}^2_{\rm L} $ and
$ {\cal M}^2_{\rm R} $ by the renormalization group effects.
For example, the flavor mixing terms are induced for the left-handed
$ d $-type squarks by the $ \lambda_u $ coupling.
The $ {\tilde d}_1 $-$ {\tilde d}_3 $ mixing term
is, in particular, calculated as
$ m^2_{{\tilde d}_1 {\tilde d}_3}
\sim m^2 t_U \lambda_t^2 V_{td}^* V_{tb} $
involving the CKM matrix.
Then, a nontrivial correction is obtained at the zero-th order of
$ \lambda_\alpha $ in the flavor changing parts of
the squark-gluino couplings.
By taking this correction together with the left-right squark coupling
term $ ( \Delta_d )_{13} \sim t_U {\bar \lambda}_D^2
( \epsilon_{\rm R} \epsilon_{\rm R}^\dagger )_{31}^* \lambda_d $
(though $ ( \Delta_d )_{13} \sim \lambda_{d_1} ( V^\dagger V )_{13}
\sim \lambda_d^3 $ in the universal limit,
as seen in eq.(\ref{eqn:dDd}) with eq.(\ref{eqn:CKM1})),
a contribution to $ d_d ({\tilde g}) $ is obtained with the factor as
\begin{equation}
t_U^2 {\bar \lambda}_D^2 \lambda_t^2 \lambda_d
{\rm Im} [ V_{td} V_{tb}^*
( \epsilon_{\rm R} \epsilon_{\rm R}^\dagger )^*_{31} ] ~,
\label{eqn:grg}
\end{equation}
where $ \epsilon_{\rm R} = \lambda_\alpha \lambda_D^{-1} $.
This contribution can also be comparable to the contribution
given in eq.(\ref{eqn:ImdDd}).

As seen in the above considerations, significant gluino contributions
to the neutron EDM are obtained from the $ d $-$ D $ coupling
$ \lambda_\alpha $ including the renormalization group effects.
These contributions,
which may generally be expressed in terms of the coupling parameters
such as $ \lambda_d $, $ \lambda_\alpha $ and the CKM matrix $ V $,
should have the forms invariant under the relevant phase transformations
of the quark fields, as given in eqs.(\ref{eqn:ImdDd}) and (\ref{eqn:grg}).
Then, they appear to be of the order of $ 10^{-26} e~{\rm cm} $
for reasonable ranges of the coupling parameters.
Detailed analyses will be presented in a forthcoming paper,
where numerical calculations are made for diagonalizing
the mass matrices and solving the renormalization group equations.

We finally comment on the case of non-universal soft supersymmetry
breakings, though it does not look appealing.
For example, if $ \xi_\beta / m $ is independent of $ \lambda_\alpha $
and $ \lambda_d $ rather than $ \lambda_\alpha^\dagger \lambda_d $
of the universal case, we obtain a gluino contribution
with $ {\rm Im}[ ( \Delta_\alpha )_{1k}( \Delta_D^\dagger )_{kl}
( \Delta_\beta )_{l1} ] $
$ \sim {\rm Im}[ ( \lambda_\alpha )_{1k} ( \xi_\beta / m )_{k1} ] $
in eq.(\ref{eqn:d-d1}).
While $ ( \lambda_\alpha )_{1k} $ can be made real,
a new phase $ \phi_\beta $ arises in $ \xi_\beta $.
This contribution is no longer proportional
to $ \lambda_d $ nor suppressed by the CKM mixing.
Then, the factor $ | \lambda_\alpha | | \xi_\beta | \phi_\beta $
is severly restricted by the experimental bound on the neutron EDM.
In this way, unnatural constraints are imposed
on the generic form of soft supersymmetry breakings,
as already seen in the MSSM.

In summary, we have investigated the $ CP $ violating effects,
especially the neutron electric dipole moment,
which are obtained from the $ d $-$ D $ coupling
in a supersymmetric model admitting
$ {\rm SU(2)}_W $ singlet and $ Q = -1/3 $ $ D $-type quarks.
The $ d $-$ D $ mixing effects on the CKM unitarity and the neutral
currents are small enough being suppressed by the ordinary $ d $-type
quark Yukawa couplings, so that the $ d $-$ D $ coupling
with new $ CP $ violating phases can take significant values.
Then, through the $ d $-$ D $ mixing effects
on the squark mass terms the gluino one-loop diagrams
provide naturally the contributions $ \sim 10^{-26} e~{\rm cm} $
comparable to the current experimental bound
on the neutron electric dipole moment.
The $ CP $ violation in the $ d $-$ D $ coupling
would also be relevant for the electroweak baryogenesis.
In fact, the quark mass eigenstates will become space-dependent
with the complex $ d $-$ D $ coupling,
since the configurations of the Higgs fields $ H_1 $, $ H_2 $ and $ N $
are in general varying in different manners through the bubble wall.
This effect would contribute significantly in determining
the local charge distributions of the quarks
to generate the baryon number asymmetry
around the walls of expanding bubbles.

\bigskip
The authors would like to thank N. Oshimo and A. Sugamoto
for valuable comments.

\newpage

\newpage
\begin{flushleft}
{\Large{\bf{Figure Caption}}}
\end{flushleft}

\begin{description}
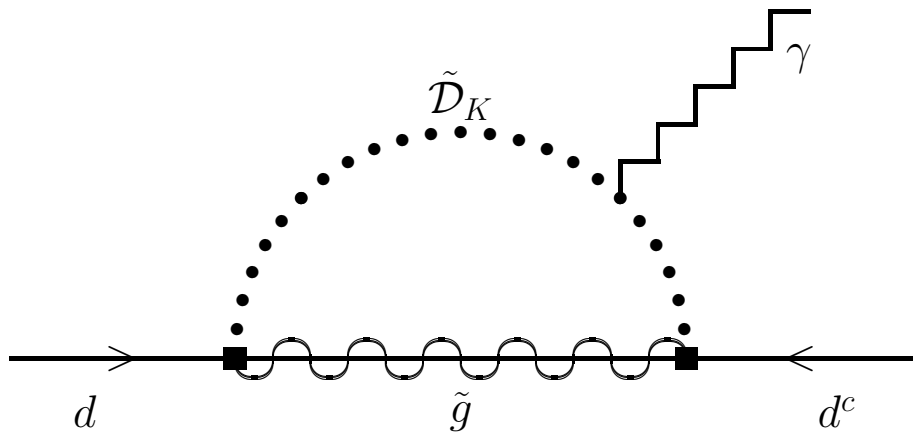
\item[Fig.1]
One-loop gluino-squark diagram contributing to the $ d $ quark EDM.
\end{description}

\newpage
\begin{figure}[h]

\unitlength 1mm

\begin{picture}(0,0)(-80,100)

\put(0,-50){\makebox(0,0){\LARGE{Fig.1}}}
{\linethickness{.5mm}
\put(-60,0){\line(1,0){120}}
\multiput(21.213,21.213)(5,5){5}{\line(0,1){5.25}}
\multiput(21.213,26.213)(5,5){5}{\line(1,0){5.25}}
}

\put(-45,0){\makebox(0,0){\Large{$ > $}}}
\put(45,0){\makebox(0,0){\Large{$ < $}}}

\put(-50,-7){\makebox(0,0){\Large{$ d $}}}
\put(50,-7){\makebox(0,0){\Large{$ d^c $}}}
\put(0,-7){\makebox(0,0){\Large{$ {\tilde g} $}}}
\put(0,35){\makebox(0,0){\Large{$ {\tilde{\cal D}}_K $}}}
\put(45,40){\makebox(0,0){\Large{$ \gamma $}}}

{\linethickness{3mm}
\put(28.5,0){\line(1,0){3}}
\put(-31.5,0){\line(1,0){3}}
}

\multiput(27.5,-.2)(-10,0){6}{\oval(5,5)[t]}
\multiput(22.5,-.2)(-10,0){6}{\oval(5,5)[b]}
\multiput(27.5,0)(-10,0){6}{\oval(5,5)[t]}
\multiput(22.5,0)(-10,0){6}{\oval(5,5)[b]}
\multiput(27.5,.2)(-10,0){6}{\oval(5,5)[t]}
\multiput(22.5,.2)(-10,0){6}{\oval(5,5)[b]}

\put(30,0){\makebox(0,0){$ \bullet $}}
\put(29.743,3.9158){\makebox(0,0){$ \bullet $}}
\put(28.978,7.7646){\makebox(0,0){$ \bullet $}}
\put(27.716,11.481){\makebox(0,0){$ \bullet $}}
\put(25.981,15.000){\makebox(0,0){$ \bullet $}}
\put(23.801,18.263){\makebox(0,0){$ \bullet $}}
\put(21.213,21.213){\makebox(0,0){$ \bullet $}}

\put(0,30){\makebox(0,0){$ \bullet $}}
\put(3.9158,29.743){\makebox(0,0){$ \bullet $}}
\put(7.7646,28.978){\makebox(0,0){$ \bullet $}}
\put(11.481,27.716){\makebox(0,0){$ \bullet $}}
\put(15.000,25.981){\makebox(0,0){$ \bullet $}}
\put(18.263,23.801){\makebox(0,0){$ \bullet $}}
\put(21.213,21.213){\makebox(0,0){$ \bullet $}}

\put(-30,0){\makebox(0,0){$ \bullet $}}
\put(-29.743,3.9158){\makebox(0,0){$ \bullet $}}
\put(-28.978,7.7646){\makebox(0,0){$ \bullet $}}
\put(-27.716,11.481){\makebox(0,0){$ \bullet $}}
\put(-25.981,15.000){\makebox(0,0){$ \bullet $}}
\put(-23.801,18.263){\makebox(0,0){$ \bullet $}}
\put(-21.213,21.213){\makebox(0,0){$ \bullet $}}

\put(-3.9158,29.743){\makebox(0,0){$ \bullet $}}
\put(-7.7646,28.978){\makebox(0,0){$ \bullet $}}
\put(-11.481,27.716){\makebox(0,0){$ \bullet $}}
\put(-15.000,25.981){\makebox(0,0){$ \bullet $}}
\put(-18.263,23.801){\makebox(0,0){$ \bullet $}}
\put(-21.213,21.213){\makebox(0,0){$ \bullet $}}

\end{picture}

\end{figure}

\end{document}